\documentclass[aps,twocolumn,showpacs]{revtex4}
\usepackage{graphics}
\usepackage{latexsym}
\begin{document}
\tighten
\bibliographystyle{apsrev}
\def\half{{1\over 2}}
\def \D {\mbox{D}}
\def\curl {\mbox{curl}\,}
\def \ep {\varepsilon}
\def \lleq {\lower0.9ex\hbox{ $\buildrel < \over \sim$} ~}
\def \ggeq {\lower0.9ex\hbox{ $\buildrel > \over \sim$} ~}
\def\beq{\begin{equation}}
\def\eeq{\end{equation}}
\def\ber{\begin{eqnarray}}
\def\eer{\end{eqnarray}}
\def \apl {ApJ, }
\def \aps {ApJS, }
\def \pd {Phys. Rev. D, }
\def \prl {Phys. Rev. Lett., }
\def \pl {Phys. Lett., }
\def \np {Nucl. Phys., }
\def \l {\Lambda}

\title{Unifying Brane World Inflation with Quintessence}
\author{M.Sami}
\altaffiliation[On leave from:]{ Department of Physics, Jamia Millia, New Delhi-110025}
\email{sami@iucaa.ernet.in}
\author{N. Dadhich}
\email{nkd@iucaa.ernet.in}
\affiliation{IUCAA, Post Bag 4, Ganeshkhind,\\
 Pune 411 007, India.}
 
\pacs{98.80.Cq,~98.80.Hw,~04.50.+h}
 
\begin{abstract}
We review the recent attempts of unifying inflation with quintessence. It appears natural to join the two ends in the framework
of brane world cosmology. The models of quintessential inflation belong to the class of {\it non-oscillatory} models for which
the mechanism of conventional reheating does not work. Reheating through gravitational particle production is inefficient and
leads to the excessive production of relic gravity waves which results in the violation of nucleosynthesis constraint. 
The mechanism of {\it instant preheating} is quite efficient
 and is suitable for brane world quintessential inflation. The model
is shown to be free from the problem of excessive production of
gravity waves. The prospects of Gauss-Bonnet brane world inflation are also briefly indicated.
\end{abstract}
\maketitle
\section{INTRODUCTION}
Universe seems to exhibit an interesting symmetry with regard to accelerated expansion. It has gone
under inflation at early epochs and is believed to be accelerating at present.
The inflationary paradigm was originally introduced to address the initial value problems of the standard hot big bang model. Only later it became clear that the scenario
could provide important clues for the origin of structure in the universe. The recent measurement of the
Wilkinson Microwave Anisotropy Probe (WMAP) \cite{WMAP1,WMAP2} in
the Cosmic Microwave Background (CMB) made it clear that (i) the
current state of the universe is very close to a critical density
and that (ii) primordial density perturbations that seeded
large-scale structure in the universe are nearly scale-invariant
and Gaussian, which are consistent with the inflationary paradigm. 
Inflation is often implemented with a single or multiple scalar-field models\cite{LR} (also see the excellent review
on inflation by Shinji Tsujikawa\cite{shinji}). In most of these models,
the scalar field undergoes a slow-roll period allowing an accelerated expansion of the universe. After drawing the required amount of inflation, the inflaton enters the regime of quasi-periodic 
oscillation where it
quickly oscillates and decays into particles leading to (p)reheating.\par
As for the current accelerating of universe, it is supported by observations of high
redshift type Ia supernovae treated as standardized candles and, more indirectly,
by observations of the cosmic microwave background and galaxy clustering.
Within the framework of general relativity, cosmic acceleration should be
sourced by an energy-momentum tensor which has a large negative
pressure (dark energy)\cite{phiindustry}.
Therefore, the standard model should, in order to comply with the logical consistency and observation, be sandwiched between inflation at early epochs and quintessence at late times. It is natural to
ask whether one can build a model with scalar fields to join the two ends without disturbing the thermal
history of universe. Attempts have been made to unify both these concepts using models with a single scalar
field \cite{unifiedmodels}.In these models, the scalar field  exhibits the properties of tracker field. As a result it goes into hiding after the commencement of radiation domination; it emerges from the shadow only at late times to account
for the observed accelerated expansion of universe. These models belong to the category of
{\it non oscillating} models in which the standard reheating mechanism does not work. In this case, one can employ an alternative mechanism of
reheating via quantum-mechanical particle production in time varying gravitational field at the end of inflation \cite{ford}. However,
then the
inflaton energy density should
red-shift faster than that of the produced particles so that radiation domination could commence. And this requires a steep
field potential, which of course, cannot support inflation in the standard FRW cosmology. This is precisely where the brane\cite{randall,h} assisted
inflation comes to the rescue.\par
The presence of the quadratic density term (high energy corrections) in the Friedman equation on the brane changes the
expansion dynamics at early epochs \cite{cline}(see Ref\cite{roy rev} for details on the dynamics of brane worlds)
Consequently, the field experiences greater damping and rolls down its potential slower than it would during the conventional inflation.
 Thus, inflation in the brane world scenario can successfully occur for very steep potentials\cite{basset,liddle}. The model of
quintessential inflation based upon reheating via gravitational particle production is faced with difficulties associated with excessive production of gravity waves. Indeed the reheating mechanism based upon this process is extremely inefficient. The energy density of so produced radiation
 is typically one part in $10^{16}$\cite{liddle}  to the scalar-field
energy density at the end of inflation. As a result, these models
have prolonged kinetic regime during which the amplitude of primordial
gravity waves enhances and violates the nucleosynthesis constraint\cite{vst}(see also \cite{star79}). Hence, it
is necessary to look for alternative mechanisms more efficient than
the gravitational particle production to address the problem. 

A proposal of reheating with Born-Infeld matter was made
in Ref\cite{bisn}(see also Ref\cite{bi1,bi2} on the related theme). It was shown that reheating is quite efficient and
the model does not require any additional fine tuning of parameters\cite{bisn}.
However, the model works under several assumptions which are not easy to justify.\par
The problems associated with reheating mechanisms discussed above can be circumvented if
one invokes an alternative method of reheating,
namely `instant preheating' proposed by Felder, Kofman and Linde
\cite{FKL} (see also Ref\cite{shtanov} on the related theme. For other approaches to reheating in quintessential inflation see
\cite{curvaton}).
This mechanism is quite efficient and  robust, and is well suited to non-oscillating models. It describes a new method of realizing quintessential inflation on the brane
in which inflation is followed by `instant preheating'. 
The larger reheating temperature in this model results in a
smaller amplitude of relic gravity waves which is consistent with the nucleosynthesis bounds\cite{samiv}. However,
the recent measurement of CMB anisotropies
by WMAP places fairly strong constraints on inflationary models
\cite{spergel03,tegmark03}.
It seems that the steep brane world inflation is on the the verge of being ruled out by the observations\cite{suji04}. 
Steep inflation in a Gauss-Bonnet braneworld may appear to be
in better agreement with observations than inflation in a RS scenario \cite{lidsey}.
\section{QUINTESSENTIAL INFLATION}
Quintessential inflation aims to describe a scenario in which both inflation and dark energy (quintessence) are described by the same scalar field. The unification of these concepts in a single scalar field model imposes certain constraints which were spelled out in the introduction. These concepts can be put together consistently in context with brane world inflation. Let us below list the building blocks of such a  model.\\
$\bullet$ {Alternative Mechanisms of Reheating}\\
(i) Reheating via gravitational particle production.\\
(2) Curvaton reheating.\\
(3) Born-Infeld induced reheating.\\
(4) Instant preheating.\\
$\bullet$ {Steep Inflaton Potential}.\\
$\bullet$ {Brane World Assisted Inflation}.\\
$\bullet$ {Tracher Field}.\\
$\bullet$ {Late Time Features in the Potential}\\
(1) Potentials which become shallow at late time (such as inverse power law potentials)\\
(2) Potentials reducing to particular power law type at late times.
\subsection{Steep Brane World Inflation}
In what follows we shall work with the steep exponential potential which exhibits the aforementioned
features necessary for the description of inflationary as well as post inflationary regimes. The
brane world inflation with steep potentials becomes possible due to high energy corrections in the Friedmann equation. The exit from inflation also takes place naturally when the high energy corrections become
unimportant.\par
In the 4+1 dimensional brane scenario
inspired by the Randall-Sundrum \cite{randall} model, the standard Friedman
equation is modified to \cite{cline}
\beq
H^2 = \frac{1}{3 M_p^2}\rho \left (1 + \frac{\rho}{2\lambda_b}\right ) + \frac{\Lambda_4}{3} +\frac{\cal E}{a^4}\label{eq:frw1}
\eeq
where ${\cal E}$ is an integration constant which transmits bulk graviton
influence onto the brane and
$\lambda_b$ is the three dimensional brane tension which provides a
relationship between the four and five-dimensional Planck masses
and also relates the four-dimensional cosmological constant $\Lambda_4$
to its five-dimensional counterpart. 

The four dimensional cosmological constant $\Lambda_4$ can be made to vanish by appropriately tuning the brane tension.
The ``dark radiation''
${\cal E}/a^4$ is expected to rapidly disappear once inflation has
commenced so that we effectively get \cite{cline,basset}
\beq
H^2 = \frac{1}{3 M_p^2}\rho \left (1 + \frac{\rho}{2\lambda_b}\right ),
\label{eq:frw2}
\eeq
where $\rho\equiv \rho_{\phi} = \half{\dot\phi}^2 + V(\phi)$, if one is dealing with a
universe dominated by a single minimally coupled scalar field.
The equation of motion of a scalar field propagating on the brane is
\begin{equation}
{\ddot \phi} + 3H {\dot \phi} + V'(\phi) = 0.
\label{eq:kg}
\end{equation}
From (\ref{eq:frw2}) and (\ref{eq:kg}) we find that the presence of the additional term
$\rho^2/\lambda_b$
increases the damping experienced by the scalar field as it rolls down its
potential. This effect is reflected in the slow-roll parameters which
have the form \cite{basset,liddle}
\ber
\epsilon &=& \epsilon_{\rm FRW}
\frac{1 + V/\lambda_b}{\left (1 + V/2\lambda_b \right )^2},\nonumber\\
\eta &=& \eta_{\rm FRW}\left (1 + V/2\lambda_b \right )^{-1},
\label{eq:slow-rollbrane}
\eer
where
\beq
\epsilon_{\rm FRW} = \frac{M_p^2}{2} \left (\frac{V'}{V}\right )^2,
\,\, \eta_{\rm FRW} = {M_p^2} \left (\frac{V''}{V}\right )
\label{eq:slow_FRW}
\eeq
are slow roll parameters in the absence of brane corrections.
The influence of the brane term becomes important when $V/\lambda_b \gg 1$
and in this case we get
\beq
\epsilon \simeq 4\epsilon_{\rm FRW} (V/\lambda_b)^{-1},\,
\eta \simeq 2\eta_{\rm FRW} (V/\lambda_b)^{-1}.
\label{eq:slow-roll}
\eeq
Clearly slow-roll ($\epsilon, \eta \ll 1$) is easier to achieve when
$V/\lambda_b \gg 1$ and on this basis one can expect inflation
to occur even for relatively steep
potentials, such the exponential and the inverse power-law which we discuss
below.\par
\subsection{Exponential Potentials}
\label{sec:exp}
 
 The exponential potential
 \beq
 V(\phi) = V_0e^{\alpha\phi/M_P}
 \label{expo}
 \eeq
 with ${\dot \phi} < 0$
 (equivalently $V(\phi) = V_0e^{-\alpha\phi/M_P}$
 with ${\dot \phi} > 0$)
 has traditionally played an important role
 within the inflationary framework  since, in the absence of matter,
 it gives rise to power law inflation $a \propto t^c$,
 $c = 2/\alpha^2$
 provided $\alpha\leq \sqrt{2}$.
 For $\alpha > \sqrt{2}$
 the potential becomes too steep to sustain inflation and for
 larger values $\alpha \geq \sqrt{6}$ the field enters a
 kinetic regime during which
field energy density $\rho_\phi \propto a^{-6}$.
 Thus within the standard general relativistic framework,
 steep potentials 
 are not capable of sustaining inflation.
 However
 extra-dimensional effects lead to interesting new possibilities for the
 inflationary scenario. The increased damping of the scalar field
 when $V/\lambda_b \gg 1$ leads to
 a decrease in the value of the slow-roll parameters
 $\epsilon = \eta \simeq 2\alpha^2\lambda_b/V$,
 so that slow-roll ($\epsilon,\eta \ll 1$) leading to inflation
 now becomes possible even for large values of $\alpha$.
 The steep exponential potentials satisfies the post inflationary requirements mentioned earlier. Infact, the cosmological dynamics with
steep exponential potential in presence of background (radiation/matter) admits scaling solution as the attractor of the system. The
attractor is characterized by the tracking behavior of the field energy density $\rho_{\phi}$. During the 'tracking regime', the ratio
of  $\rho_{\phi}$ to the background energy density $\rho_B$ is held fixed
\begin{equation}
{\rho_{\phi} \over {\rho_{\phi}+\rho_B}}={{3\left (1+w_B \right)} \over \alpha^2}\lleq 0.2
\label{track}
\end{equation}
where $w_B$ is the equation of state parameter for background ($w_B=0,~1/3$ for matter and radiation respectively) and the inequality (\ref{track}) reflects the nucleosynthesis constraint which requires
$\alpha \ggeq 5$. It is therefore clear that the field energy density in the post inflationary regime would keep tracking the
background being subdominant such that it does not interfere with the thermal history of the universe. 
 
  Within the framework of the braneworld scenario, the field equations
  (\ref{eq:frw2}) and (\ref{eq:kg}) can be solved {\em exactly} in the
  slow-roll limit when
  $\rho/\lambda_b \gg 1$. In this case
  \beq
  \frac{\dot{a}(t)}{a(t)} \simeq \frac{1}{\sqrt{6 M_P^2\lambda_b}} V(\phi),
  \label{eq:frw3}
  \eeq
  which, when substituted in
  \beq
  3H\dot{\phi} \simeq -V'(\phi)
  \label{eq:kg1}
  \eeq
  leads to 
\begin{equation}
{\dot\phi}(t) = - \alpha\sqrt{2\lambda_b/3}
\label{phidot}
\end{equation}
  The expression for number of inflationary e-foldings is easy to establish
  \ber
  {\cal N} &=& \log{\frac{a(t)}{a_i}} = \int_{t_i}^t H(t') dt'\\
           &=& \nonumber \frac{V_0}{2\lambda_b\alpha^2}(e^{\alpha\phi_i} - e^{\alpha\phi(t)})\label{eq:efold0}\\
  &=& \frac{V_i}{2\lambda_b \alpha^2}\left [ 1 -
  \exp{\lbrace-\sqrt{\frac{2\lambda_b}{3M_P^2}\alpha^2}(t-t_i)\rbrace}\right ],
  \label{eq:efold}
  \eer
  where $V_i = V_0e^{\alpha\phi_i}$.
  From Eq.~(\ref{eq:efold}) we find that the expansion factor passes
  through an inflection point marking the end of inflation and leading to 
  \beq
  \phi_{\rm end} =
  -\frac{M_P}{\alpha}
  \log{\bigg (\frac{V_0}{2\lambda_b \alpha^2}\bigg )},
  \label{phiend}
  \eeq
  \beq
  V_{\rm end} \equiv V_0e^{\alpha\phi_{\rm end}/M_P} =
  2\lambda_b\ \alpha^2\\
  \label{eq:Vend}.
  \eeq
   The COBE normalized value for the amplitude of scalar density perturbations allows to estimate $V_{end}$ and the brane tension
   $\lambda_b$
   \ber
V_{end} &\simeq& {{3 \times 10^{-7}} \over \alpha^4}
\left({M_p \over {\cal N }+1}\right)^4
\nonumber\\
\lambda_b &\simeq& {{1.3 \times 10^{-7}} \over \alpha^6}
\left({M_p \over {\cal N }+1}\right)^4   ~,
\label{vend}
\eer 
   We work here under the assumption that scalar density
   perturbations are responsible for most of the COBE signal. We shall, however, come 
back to the important question about the tensor perturbations later in our discussion.\par
The scenario of quintessential inflation we are discussing here belongs to the class of
non-oscillatory models where the conventional reheating mechanism does not work. We can use
the gravitational particle production to do the required. This is a democratic process which leads
to the production of a variety of species quantum mechanically at the end of inflation when
the space time geometry suffers a crucial change. Unlike the conventional reheating mechanism,
this process does not requite the introduction of extra fields. The radiation density created
via this mechanism at the end of inflation is given by
\begin{equation}
\rho_r \sim 0.01 \times g_p H_{end}^4 
\label{grad}
\end{equation}
where $g_p \sim 100$ is the number of different particle species created from vacuum. Using the relation (\ref{grad}) and the expressions of $\lambda_b$ and $V_{end}$ obtained above, it can easily be shown that
\beq
\left (\frac{\rho_\phi}{\rho_r}\right)_{\rm end}  \sim 2 \times 10^{16}\left (\frac{{\cal N} + 1}{51}\right )^4
g_p^{-1}
\label{eq:ratio_end}
\eeq

This leads to a prolonged `kinetic regime' during which scalar matter has
the `stiff' equation of state .

     Using Eqs.~(\ref{eq:efold0}) \& (\ref{eq:efold}) 
     one can demonstrate
     that inflation proceeds at an exponential rate during early epochs which plays an important role
     for the generation of relic gravity waves during inflation.
\section{Late Time Evolution}
\begin{figure}[h]
\centering
\resizebox{!}{2.5in}{\includegraphics{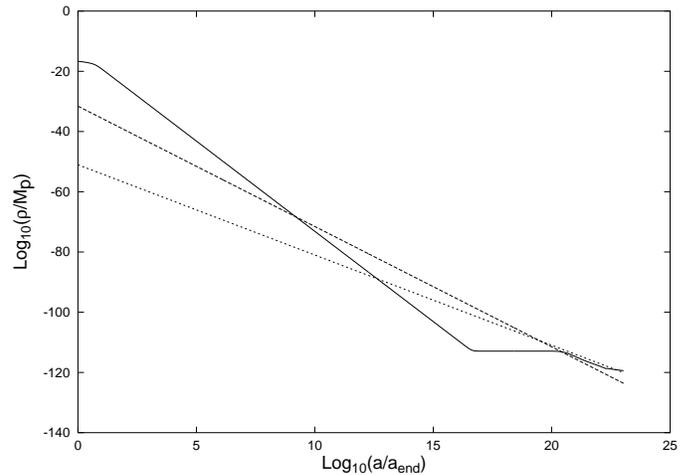}}
\caption{The post-inflationary energy density in
the scalar field (solid line) radiation (dashed line) and cold dark
matter (dotted line)
is shown as a function of the scale factor for the model
described by (\ref{cosine}) with
$V_0 \simeq 5\times 10^{-46}$ GeV$^4$, $\tilde\alpha = 5$ and $p = 0.2$.
The enormously large value of the scalar field
kinetic energy (relative to the potential)
ensures that the scalar field density overshoots the background
radiation value, after which $\rho_\phi$ remains approximately constant
for a substantially long period of time. At late times the scalar field briefly
tracks the background matter density before becoming dominant
and driving the current accelerated expansion of the universe. From Sahni, Sami and Souradeep\cite{vst}
}
\label{fig:cosh1}
\end{figure}

\begin{figure}[h]
\centering
\resizebox{!}{2.5in}{\includegraphics{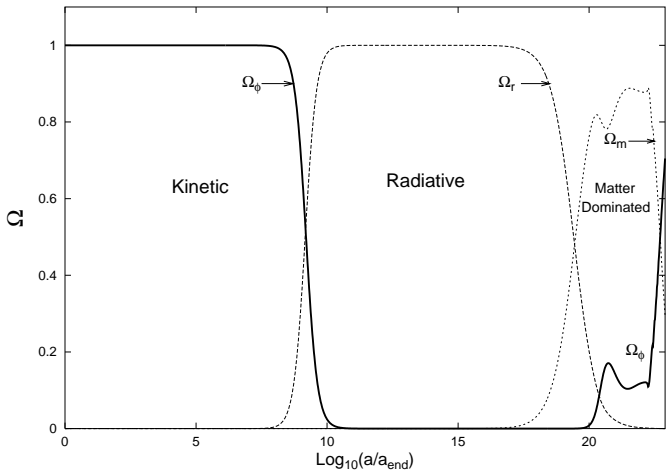}}
\caption{The dimensionless density parameter $\Omega$ is plotted as
a function of the scale factor for the model in figure \ref{fig:cosh1}.
Late time oscillations of the scalar field ensure that
the mean equation of state turns negative
$\langle w_\phi\rangle \simeq -2/3$, giving rise to the current
epoch of cosmic acceleration with $a(t) \propto t^2$ and present day values
$\Omega_{0\phi} \simeq 0.7, \Omega_{0m} \simeq 0.3$. From Sahni, Sami and Souradeep\cite{vst}}
\label{fig:cosh2}
\end{figure}

As discussed above, the scalar field with exponential potential (\ref{exppot}) leads to a viable
evolution at early times. We should, however, ensure that the scalar field becomes quintessence at late times 
which demands a particular behavior of the scalar field potential as discussed above. Indeed, 
any scalar field potential which interpolates between an exponential at early epochs and the power law type potential at late times could lead to a viable cosmological evolution.
The {\it cosine
hyperbolic} potential  provides one such example
\cite{sahni}

\begin{equation}
V(\phi)=V_0\left[\cosh (\tilde{\alpha}\phi/M_p)-1\right]^p,~~~~p~>0
\label{cosine}
\end{equation}
which has asymptotic forms
\begin{equation}
V(\phi)={V_0 \over 2^p} e^{{\alpha}\phi/M_p},~~~\tilde{\alpha}\phi/M_p~>>1,~~\phi~>0
\label{exppot}
\end{equation}
\begin{equation}
V(\phi)={V_0 \over 2^p}\left(\tilde{\alpha} \phi \over M_p \right)^{2p}~~~~~~|\tilde{\alpha} \phi/M_p|~<<1
\end{equation}
where ${\alpha}=p\tilde{\alpha}$. 
As the {\it cosine
hyperbolic} potential (\ref{cosine}) exhibits power law type of behavior near the origin, field oscillations build up in the system
at late times. For a particular choice of power law, the average equation of state parameter may turn negative\cite{sahni,turner}
\begin{equation}
\left< w_{\phi} \right> = \left <{{ {\dot{\phi}^2 \over 2}-V(\phi)} \over {{\dot{\phi}^2 \over 2}+V(\phi)}} \right >={{p-1} \over{p+1}}
\label{stateeq}
\end{equation}
As a result the scalar field energy density and the scale factor have the following behavior
$$\rho_{\phi} \propto  a^{-3(1+\left< w \right>)},~~~~~~~~a \propto   t^{{ 2 \over 3}(1 + \left< w \right>  )^{-1}}. $$
The average equation of state
$ \left<w(\phi) \right> <-1/3$ for $p < 1/2$ allowing the scalar field to play the role of dark energy.We have numerically solved for the behaviour of this model after
including a radiative term (arising from inflationary particle
production discussed in the previous section) and standard cold dark matter.
Our results for a particular realization of the model 
are shown in figures \ref{fig:cosh1} \& \ref{fig:cosh2}.
We find that, due to the very large value of the scalar field kinetic energy
at the commencement of the radiative regime,
the scalar field density
overshoots the radiation energy density.
After this, the value
of $\rho_\phi$ stabilizes and
remains relatively unchanged
for a considerable length of time during which the scalar field
equation of state is $w_\phi \simeq -1$.
Tracking commences late into
the matter dominated epoch and the universe accelerates today during rapid
oscillations of the scalar field.
This model provides an interesting example of `quintessential
inflation'. However as we shall discuss next,
the long duration of the kinetic regime in this model results in a
large gravity wave background which comes into conflict with
nucleosynthesis constraints.
 
     \subsection{Relic Gravity Waves and Nucleosynthesis Constraint}
\label{gravity}
The tensor perturbations or gravity waves get quantum mechanically generated during inflation and leave imprints
on the micro-wave background. 

Gravity waves in a spatially homogeneous and isotropic background geometry
satisfy the minimally coupled Klein-Gordon equation $\Box  h_{ik} = 0$ ,
which, after a separation of variables
$h_{ij} = \phi_k(\tau) e^{-i{\bf k}{\bf x}}e_{ij}$
($e_{ij}$ is  the polarization tensor) reduces to
\begin{equation}
{\ddot \phi_k} + 2\frac{\dot a}{a}{\dot \phi}_k + k^2\phi_k = 0
\label{eq:kg2}
\end{equation}
where $\tau = \int dt/a(t)$ is the conformal time coordinate
and $k = 2\pi a/\lambda$ is the comoving wavenumber.
Since brane driven inflation is near-exponential we can write
$a = \tau_0/\tau$ $(\vert\tau\vert < \vert\tau_0\vert)$, in this case
normalized positive frequency solutions of (\ref{eq:kg2})
corresponding to the adiabatic
vacuum in the `in state' are given by \cite{vst}
\begin{equation}
\phi_{\rm in}^+(k,\tau) = \left (\frac{\pi \tau_0}{4}\right )^{1/2}
\left (\frac{\tau}{\tau_0}\right )^{3/2}
H_{3/2}^{(2)}(k\tau)F(H_{\rm in}/\tilde\mu)
\label{eq:hankel}
\end{equation}
where $\tilde\mu = M_5^3/M_p^2$ and $H_{\rm in} \equiv -1/\tau_0$ is the
inflationary Hubble parameter.
The term
\begin{equation}
F(x) = \left (\sqrt{1+x^2} - x^2\log{\lbrace \frac{1}{x} + \sqrt{1 + \frac{1}{x^2}}\rbrace }
\right )^{-1/2}
\end{equation}
is responsible for the increased gravity wave amplitude
in braneworld inflation \cite{lmaartans}.The `out state' is described by a linear superposition of positive and negative
frequency solutions to (\ref{eq:kg2}). For power law expansion
$a = (t/t_0)^p \equiv (\tau/\tau_0)^{1/2-\mu}$, we have
\begin{equation}
\phi_{\rm out}(k,\tau) = \alpha \phi^{(+)}_{\rm out}(k\tau) +
\beta \phi^{(-)}_{\rm out}(k\tau)
\label{eq:hankel1}
\end{equation}
The energy density of relic gravity waves is given by \cite{vst}
\begin{equation}
\rho_{\rm g} = \langle T_0^0\rangle =\frac{1}{\pi^2 a^4}\int
dk k^3 \vert\beta\vert^2,
\label{eq:gw_energy}
\end{equation}

Computing the Bogolyubov coefficient $\beta$, one can show that the spectral energy
density of gravity waves produced during slow-roll inflation is\cite{vst,samiv} 
\beq
\rho_g(k) \propto k^{2\left (\frac{w - 1/3}{w + 1/3}\right )}~.
\eeq
where $w$ is the equation of state parameter which characterizes the post-inflationary epoch.
In braneworld model under consideration,
$w \simeq 1$ during the kinetic regime, consequently the gravity wave background
generated during this epoch will have a
blue spectrum $\rho_g(k) \propto k$.\par
We imagine that radiation through some mechanism was generated at the end of inflation with radiation density $\rho_r$. Then the ratio of energy in gravity waves to $\rho_r$
at the commencement of
radiative regime is given by\cite{vst}
\begin{equation}
\left({\rho_g \over \rho_{r}} \right)_{eq}={64 \over {3 \pi}} h_{GW}^2 \left({T_{kin} \over T_{eq}} \right)^2
\label{graden}
\end{equation}
where $h_{GW}$ is the dimensionless amplitude of gravity waves (from COBE normalization,~
$h_{GW}^2 \simeq 1.7\times 10^{-10}$,~for ${\cal N} \simeq 70$).
We should mention that the commencement of the kinetic regime is not instaneous and the brane effects petrsist for some time after inflation has ended. The temperature at the commencement of the kinetic regime $T_{kin}$ is related to the temperature at the end of inflation as
\begin{equation}
T_{kin}=T_{end}\left(a_{end} \over a_{kin} \right) = T_{end} F_1(\alpha)
\label{Tkin}
\end{equation}
where $F_1(\alpha)=\left(c+{d\over \alpha^2} \right)$, $c \simeq 0.142$,~ $d \simeq -1.057$ and
$T_{end}=\left(\rho_{r}^{end} \right)^{1/4}$. The equality between scalar field matter and radiation
takes place at the temperature
\begin{equation}
T_{eq}= T_{end} {F_2(\alpha) \over {\left(\rho_{\phi}/\rho_{r}\right)_{end}^{1/2}}}
\label{Teq}
\end{equation}
with $ F_2(\alpha)=\left(e+{f \over \alpha^2} \right)$, $e\simeq 0.0265$,~$f\simeq -0.176$. The
fitting formulas (\ref{Tkin}) and (\ref{Teq}) are obtained by numerical integration of equations of motion.
 
Using equations (\ref{Tkin}), (\ref{Teq}) and (\ref{graden}) we obtain the ratio of scalar field energy density to radiation ener density at the end of inflation
\begin{equation}
{\left(\rho_{\phi} \over \rho_{r}\right)_{end}}=
{3 \pi \over  64 }\left({1 \over {h_{GW}^2\left(F_1(\alpha)/F_2(\alpha)\right)^2}}\right)\left({\rho_g \over \rho_{r}} \right)_{eq}
\label{endratio}
\end{equation}
Equation (\ref{endratio} ) is an important result which sets a limit on the ratio of scalar field energy density to radiation energy
density at the end of inflation. Indeed, For the nucleo-synthesis constraint to be respected, the ratio of energy density in gravity waves to radiation energy density at equality $(\rho_g/\rho_{r})_{eq} \lleq 0.2$. 
For a generic steep exponential potential (($\alpha\ggeq 5$), we have 
\begin{equation}
\left(\rho_{\phi}/\rho_{r}\right)_{end} \lleq 10^7
\label{endration} 
\end{equation}
As emphasized earlier, this ratio is of the order of $10^{16}$ in case gravitational particle production and
exceeds the nucleo-synthesis constraint by nine orders of magnitudes. 
An interesting proposal which can circumvent this difficulty has recently been suggested by Liddle and Lopez\cite{curvaton}.
The authors have employed
a new method of reheating via curvaton to address the problems associated with gravitational particle
production mechanism. The curvaton model as shown in Ref\cite{curvaton} can in principal resolve the difficulties related to excessive amplitude of
short-scale gravitational waves. Although this model is interesting, it operates through a very complex network of constraints dictated by the fine tuning
of parameters of the model.\par

In the following section, we shall examine an alternative mechanism based upon Born-Infeld reheating.
 \section{Born-Infeld Brane Worlds}
The D-branes are fundamental objects in string theory. The end points of the open string to which the gauge fields are attached are
constrained to lie on the branes. As the string theory contains gravity, the D-branes are the dynamical objects. The effective
D-brane action is given by the Born-Infeld action
\begin{equation}
S_{BI}=-\lambda_b \int{d^4x \sqrt{-{\rm det}\left(g_{\mu \nu}+F_{\mu \nu}\right)}}
\label{bi}
\end{equation}
where $F_{\mu \nu}$ is the elecromagnetic field tensor (Non-Abelian gauge fields could also be included in the action) and
$\lambda_b$ is the brane tension. The Born-Infeld action, in general, also includes Fermi fields and scalars which have been dropped here for simplicity. In the brane
world scenario {\it a la} Randall-Sundrum one adopts the Nambu-Goto action instead of the Born-Infeld action. Shiromizu 
{\it et al} have suggested that in the true spirit of the string theory, the total action in the brane world cosmology be composed of the bulk and D-brane
actions\cite{bi1}
\begin{equation}
S=S_{bulk}+S_{BI},
\end{equation}
 where $S_{bulk}$ is the five dimensional Einstein-Hilbert action with the negative
cosmological constant.
The stress tensor appearing on the right hand side (RHS) of the Einstein equations on the brane will now
be sourced by the Born-Infeld
action. The modified Friedman equation on a spatially flat FRW brane acquires the form
\begin{equation}
H^2={1 \over 3M_p^2} \rho_{\rm BI} \left(1+{\rho_{\rm BI} \over 2\lambda_b} \right)
\label{friedman}
\end{equation}
with $\rho_{\rm BI}$ given by
\begin{equation}
\rho_{\rm BI}=\epsilon+{\epsilon^2 \over {6\lambda_b}}
\end{equation}
where $E^2=B^2=\epsilon$
The tension $\lambda_b$ is tuned so that the
net cosmological constant on the brane vanishes.
We have dropped the `dark radiation' term in the equation
 (\ref{friedman}) as it rapidly disappear once inflation
sets in. Spatial averaging is assumed while computing $\rho_{\rm BI}$
and $P_{BI}$ from the stress-tensor
corresponding to action (\ref{bi}). The scaling of energy density of the
Born-Infeld matter, as usual, can be established from the conservation equation Born-Infeld matter, as usual, can be established from the conservation equation
\begin{equation}
\dot{\rho}_{\rm BI}+3H(\rho_{\rm BI}+P_{\rm BI})=0
\end{equation}
where
\begin{equation}
P_{BI}={\epsilon \over 3}-{\epsilon^2 \over {6\lambda_b}}
\label{pressureeq}
\end{equation}
Interestingly, the pressure due to the Born-Infeld matter becomes negative in the high energy regime allowing the accelerated expansion
at early times without the introduction of a scalar field. As shown in \cite{bi1} , the energy density $\rho_{\rm BI}$ scales as radiation when
$\epsilon << 6\lambda_b$. For $\epsilon> 6\lambda_b$, the Born-Infeld matter energy density starts scaling slowly (logarithmically)
with the scale factor to mimic the cosmological constant like behavior. The point is that the Born-Infeld 
matter is subdominant during  the inflationary stage. It comes to play the important role after the end of inflation
when it behaves like radiation and hence serves as an alternative to reheating mechanism.\par     

The brane world cosmology based upon the Born-Infeld action looks promising as it is perfectly tuned with the D-brane ideology. But
since the Born-Infeld action is composed of the non-linear elecromagnetic field, the D-brane cosmology proposed in
Ref\cite{bi1} can not accommodate
density perturbations at least in its present formulation. One could include a scalar field in the Born-Infeld action, say, a tachyon condensate to correct the situation. However, such a scenario faces the difficulties associated with reheating\cite{reh,linde} and formation of
acoustics/kinks\cite{kink}. We shall therefore not follow this track. We shall assume that the scalar field driving the inflation (quintessence) on the
brane is described by the usual four dimensional action for the scalar fields. We should remark here that the problems faced by rolling tachyon models are beautifully circumvented in the scenario 
based upon massive Born-Infeld scalar
field on the $\bar{D}_3$ brane of KKLT vacua\cite{garousi}.\par
\begin{figure}
\vspace{0.5in}
\resizebox{3.5in}{!}{\includegraphics{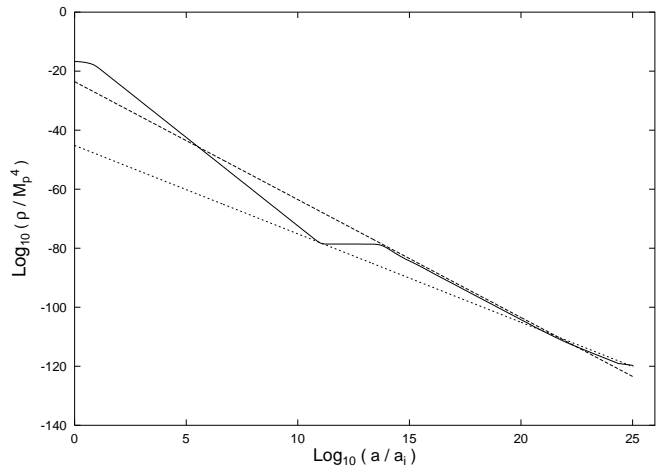}}
\caption{The post-inflationary evolution of the scalar field energy density
(solid line),
radiation (dashed line) and cold dark matter (dotted line) is shown as a
function of the scale factor for the quintessential inflation model described by
(\ref{cosine})
with $V_0^{1/4} \simeq 10^{-30} M_p$,
$\tilde{\alpha}=50$ and $p=0.1$ ($\alpha=p \tilde{\alpha}=5$).
After brane effects have ended, the field energy density $\rho_{\phi}$ enters the
kinetic regime and soon drops below the radiation density.
After a brief interval during which $<w_\phi> \simeq -1$, the scalar field begins
to track first radiation and then matter.
At very late times (present epoch) the scalar field plays the role of quintessence
and makes the universe accelerate.
The evolution of the energy density
is shown from the end of inflation until the present epoch. From Sami, Dadhich and Shiromizu\cite{bisn}
}
\label{bden}
\end{figure}

  
The total action that we are trying to motivate here is given by
\begin{equation}
S=S_{bulk}+S_{BI}+S_{\rm 4d-scalar}
\end{equation}
where
\begin{equation}
S_{\rm 4d-scalar}=-\int{\left({1 \over 2}g^{\mu \nu} \partial_{\mu} \phi \partial_{\nu} \phi+V(\phi)\right) \sqrt{-g}d^4x}
\label{NG}
\end{equation}
The energy momentum tensor for the field $\phi$ which arises from the action (\ref{NG}) is given by
\begin{equation}
T_{\mu \nu}=\partial_{\mu} \phi \partial_{\nu} \phi-g_{\mu \nu} \left[{1 \over 2}g^{\mu \nu} \partial_{\mu} \phi \partial_{\nu} \phi+V(\phi)\right]
\end{equation}
The scalar field propagating on the brane modifies the Friedman equation to
\begin{equation}
H^2={1 \over 3M_p^2} \rho_{\rm tot} \left(1+{\rho_{tot} \over 2\lambda_b} \right)
\label{mfriedman}
\end{equation}
where $\rho_{tot}$ is given by
\begin{equation}
\rho_{tot}=\rho_{\phi}+\rho_{\rm BI}
\end{equation}
As mentioned earlier, in the scenario based upon reheating via quantum mechanical
particle production during inflation, the radiation density is very small, typically one part in $10^{16}$ and the ratio of the field
energy density to that of radiation has no free parameter to tune. This leads to long kinetic regime which results in an unacceptably
large gravity background. The Born-Infeld matter which behaves like radiation (at the end
of inflation) has no such problem and can be used for reheating
without conflicting with the nucleosynthesis constraint. Indeed, at the end of inflation $\rho_{\rm BI}$ can be chosen such that $\rho_{\rm BI}^{end}<<6\lambda_b$. Such an initial condition
for $\rho_{\rm BI}$ is consistent with the nucleo-synthesis constraint\cite{bisn}. In that
case the Born-Infeld matter energy density would scale like radiation at the end of inflation. At this epoch the scale factor will be initialized at
$a_{end}=1$. The energy density $\rho_{\rm BI}$ would continue scaling as $1/a^4$ below $a=a_{end}$. The scaling would slow down as $\rho_{\rm BI}$
reaches $6\lambda_b$ which is much smaller than $V_{end}$ for generic steep potentials, say  for $\alpha\ge 5$. Hence 
$\rho_{\rm BI}$ remains
subdominant to scalar field energy density $\rho_{\phi}$ for the entire inflationary evolution. The Born-Infeld matter comes to play the
important role only at the end of inflation which is in a sense similar to curvaton. But unlike curvaton, it does not
contain any new parameter. The numerical results for a specific choice of parameters is shown in
Fig.~\ref{bden}. In contrast to the `quintessential
inflation' based upon the gravitational particle production mechanism where the scalar field spends long time in the kinetic regime and makes deep undershoot followed by long locking period with very brief tracking, the scalar field in the present scenario tracks
the background for a very long time (see figure \ref{bden}). This pattern of evolution is consistent with the
thermal history of the universe. We note that
`quintessential inflation' can also be implemented by inverse power law potentials. Unfortunately, 
one has to make several assumptions to make the scenario working:~ 
i) The tension
of the D3 brane appearing in the Born-Infeld action is treated as constant and is identified with the brane tension in the Randall-Sundrum
scenario. (ii) The fluctuations in the  Born-Infeld matter are neglected.
(iii)The series expansion of the Born-Infeld action is truncated beyond
a certain order.
In what follows, we shall examine the instant reheating mechanism
discovered by Felder, Kofman and Linde and show that their mechanism is superior to
other reheating mechanism mentioned above. 

\section{Braneworld Inflation Followed by Instant Preheating}
 
Braneworld Inflation induced by the steep exponential potential (\ref{expo}) ends when
$\phi=\phi_{end}$, see (\ref{phiend}).
Without loss of generality, we can make the inflation end at the
origin by translating the field
\begin{equation}
V(\phi') \equiv V(\phi)=\tilde{ V_0} e^{\alpha \phi'/M_p}~,
\label{texp}
\end{equation}
where $\tilde{ V_0}=V_0e^{+\alpha \phi_{end}}/M_p $ and $\phi'=\phi-\phi_{end}$.
In order to achieve reheating after inflation has ended we assume that the inflaton
$\phi$ interacts with
another scalar field $\chi$ which has a Yukawa-type interaction with
a Fermi field $\psi$. The interaction Lagrangian is
 
\begin{figure}
\resizebox{3.0in}{!}{\includegraphics{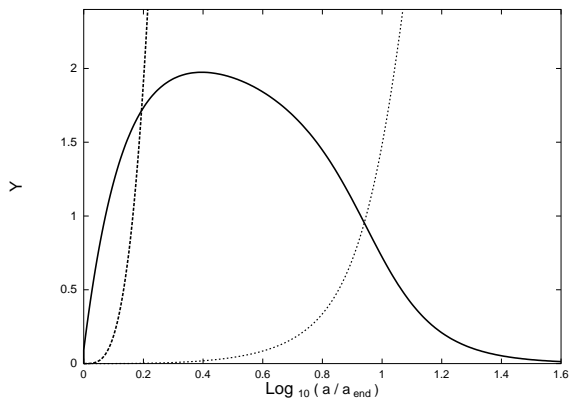}}
\caption{ The post inflationary evolution of $Y=10^9 \times |\dot{\phi}|/M_p^2 $,~ $g \phi^2/M_p^2$ is shown as a function of the scale factor for the model described by
Eq (\ref{cosine}) with $V_0\simeq 10^{-30} M_p$,
$\tilde{\alpha}=50$ and $p=0.1$.
The solid curve corresponds to $\dot{\phi}/M_p^2$ whereas the dashed and dotted curves
correspond to
$ g \phi^2/M_p^2$ for $g=10^{-6}$ (extreme left) and $g=10^{-9}$ (extreme right)
respectively. The violation of the adiabaticity condition (necessary
for particle production to take place) arises in the region bounded by the solid and dashed curves. The time duration of
particle production is seen to be smaller for larger values of the coupling $g$.
For the range of $g$
allowed by
the nucleosynthesis constraint, particle production is almost instantaneous. From Sami and Sahni\cite{samiv}}
\label{nadiabatic}
\end{figure}
 
\begin{equation}
L_{int}=-{1\over 2}g^2 \phi'^2 \chi^2-h \bar{\psi}\psi \chi~.
\label{lagrangian}
\end{equation}
To avoid confusion, we drop the prime on $\phi$ remembering that $\phi < 0$
after inflation has ended.
It should be noticed that the $\chi$ field has no bare mass,
its effective mass being determined by the field $\phi$ and the value of the coupling
constant $g$ ($m_{\chi}=g\vert\phi\vert$).
 
The production of $\chi$ particles commences as soon as $m_{\chi}$ begins changing
non-adiabatically \cite{FKL}
\begin{equation}
|\dot{m_{\chi}}| \ggeq m_{\chi}^2~~~or~~~~|\dot{\phi}| \ggeq g\phi^2~.
\label{ad}
\end{equation}
The condition for particle production (\ref{ad}) is satisfied when
\begin{equation}
\vert \phi\vert \lleq \vert \phi_{prod}\vert
= \sqrt{\frac{\vert\dot{\phi}_{end}\vert}{g}}
=\sqrt{{V_{end}^{1/2} \over {\sqrt{3} g}}}~.
\label{phiprod}
\end{equation}
From (\ref{vend}) we find that $\phi_{prod}\ll M_p$ for $g\gg 10^{-9}$.
The production time for $\chi$ particles can
be estimated to be
\begin{equation}
\Delta t_{prod} \sim {\vert \phi\vert \over {|\dot{\phi}|}}
\sim \frac{1}{\sqrt{g V_{end}^{1/2}}}~.
\label{tprod}
\end{equation}
The uncertainty relation provides an estimate for the momentum of $\chi$ particles
created non-adiabatically: $k_{prod} \simeq (\Delta t_{prod})^{-1}
\sim g^{1/2} V_{end}^{1/4}$. Proceeding as in \cite{FKL} we can show that
the occupation number of $\chi$ particles jumps sharply from zero to
\begin{equation}
n_k \simeq \exp(-{\pi k^2/{g V_{end}}^{1/2}})~,
\end{equation}
during the time interval $\Delta t_{prod}$.
The $\chi$-particle number density is estimated to be
\begin{equation}
n_{\chi}={1 \over {2\pi^3}} \int_0^{\infty}{k^2n_k} dk
\simeq {(gV_{end}^{1/2})^{3/2} \over {8\pi^3}}~.
\label{nchi}
\end{equation}
Quanta of the $\chi$-field are created during the time interval $\Delta t_{prod}$
that the field $\phi$ spends in the vicinity of $\phi = 0$. Thereafter the mass
of the $\chi$-particle begins to grow since $m_\chi = g\vert\phi(t)\vert$, and
the energy density of particles of the $\chi$-field created in this manner is
given by
\begin{equation}
\rho_{\chi} = m_\chi n_{\chi}\left({a_{end} \over a}\right)^3 =
{(gV^{1/2}_{end})^{3/2} \over
{8\pi^3}} {{g \vert\phi(t)\vert}}\left({a_{end} \over a}\right)^3~.
\end{equation}
where the $({a_{end} / a})^3$ term accounts for the cosmological
dilution of the energy density with time.
As shown above, the process of $\chi$ particle-production
takes place immediately after inflation has ended, provided $g \ggeq 10^{-9}$.
In what follows we will show that the $\chi$-field can rapidly decay into fermions.
It is easy to show that if the quanta of the $\chi$-field were converted (thermalized)
into radiation instantaneously, the radiation energy density would become 
\begin{equation}
\rho_r \simeq \rho_\chi \sim  {(gV^{1/2}_{end})^{3/2} \over {8\pi^3}} g \phi_{prod} \sim
10^{-2} g^2 V_{end}~.
\label{irad}
\end{equation}
From equation (\ref{irad}) follows the important result
\begin{equation}
\left({\rho_\phi  \over \rho_r}\right)_{end}  \sim \left (\frac{10}{g}\right )^2~.
\label{irad1}
\end{equation}

Comparing (\ref{irad1}) with (\ref{endration}) we find that, in order for
relic gravity waves to respect the nucleosynthesis constraint, we should
have $g \ggeq 4\times 10^{-3}$.
(The energy density created by instant preheating
$\left (\rho_r/\rho_\phi\right ) \simeq (g/10)^2$ can clearly be
much larger than the energy density
produced by quantum particle production, for which
$\left (\rho_r/\rho_\phi\right ) \simeq 10^{-16}g_p$.)
The constraint
$g \ggeq 4\times 10^{-3}$, implies that the particle production time-scale
(\ref{tprod}) is much smaller than the Hubble time since
\beq
\frac{1}{\Delta t_{prod} H_{end}} \ggeq 300 \alpha^2, ~~\alpha \gg 1~.
\eeq
Thus the effects of expansion can safely be neglected
during the very short time interval in which
`instant preheating' takes place.
We also find, from equation (\ref{phiprod}), that $\vert \phi_{prod}\vert
/M_p \lleq 10^{-3}$
implying that particle production takes place in a very narrow band around
$\phi = 0$.
Figure \ref{nadiabatic} demonstrates
the violation of the adiabaticity condition (at the end of inflation) which is a
necessary prerequisite
for particle production to take place.
For the range of $g$
allowed by
the nucleosynthesis constraint, the particle production turns out to be almost instantaneous.\par
We now briefly mention about the back-reaction of created $\chi$-particles
on the background. As shown in Ref\cite{samiv}, for any generic value of the coupling $g \lleq 0.3$, the back-reaction
of $\chi$ particles in the evolution equation is negligible during the time
scale $\sim H_{kin}^{-1}$ ( Here $\sim H_{kin}^{-1}$ characterizes the epoch the kinetic regime commences. $H_{kin}=H_{end}(0.085-0.688/\alpha^2)$\cite{vst}).

We now turn to the matter of reheating which occurs through the
decay of $\chi$ particles to fermions, as a consequence of the interaction term in the
Lagrangian (\ref{lagrangian}).
The decay rate of $\chi$ particles is given by
$\Gamma_{\bar{\psi}\psi}=h^2m_{\chi}/8\pi$, where $m_{\chi} = g\vert\phi\vert$.
Clearly the decay rate is faster for larger values of $\vert\phi\vert$.
For $\Gamma_{\bar{\psi}\psi} >H_{kin}$, the decay process will
be completed within the time that back-reaction effects (of $\chi$ particles)
remain small. Using the expression for ${Hkin}$ this requirement translates into
\begin{equation}
h^2 > {{8\pi \alpha} \over {\sqrt{3} g\phi}} {V_{end}^{1/2} \over M_p}F(\alpha)~.
\label{Gamma}
\end{equation}
For reheating to be completed by $\phi/M_p \lleq 1$,
we find  from  equations (\ref{endration}) and
 (\ref{Gamma})  that $ h \ggeq 10^{-4} g^{-1/2}$($g \ggeq 4 \times 10^{-3}$) for $\alpha \simeq 5$. This along with the
constraint imposed by the back reaction defines the allowed region in the parameter space (g, h). 
We observe that there is a wide region in the parameter space for which (i) reheating is  rapid and
(ii) the relic gravity background in non-oscillatory braneworld models of
quintessential inflation
is consistent with nucleo-synthesis constraints. However, this is not the complete story. One should
further subject the model to the recent WMAP observations. The measurement of CMB anisotropies
places fairly strong constraints on inflationary models
\cite{spergel03,tegmark03}.
It appears that the tensor perturbations are not adequately suppressed in the models of steep brane world inflation 
and as a result these models are on the verge of being ruled out.
As indicated by Lidsey and Nunes, inflation in a Gauss-Bonnet braneworld could appear to be
in better agreement with observations than inflation in a RS II scenario \cite{lidsey}.
In the following section, we briefly discuss the prospects of brane world inflation with the 
Gauss-Bonnet correction term in the bulk.

\section{Gauss-Bonnet Brane Worlds}
Though we are trying to motivate the GB term in the bulk having a specific
application in mind, the Gauss-Bonnet
correction is interesting in its own right and has a deep meaning. Let us
begin at the very beginning and ask for the compelling physical motivation
for general relativity (GR). It is the interaction of zero mass particle
with gravitation. Zero mass particle has the universal constant speed
which can not change yet it must feel gravity. This can only happen if
gravitational field curves space. Since space and time are already bound
together by incorporation of zero mass particle in mechanics,
gravitational field thus curves spacetime. In other words it can truly be
described by curvature of spacetime and it thus becomes a property of
spacetime - no longer an external field \cite{n1}\par
From the physical standpoint the new feature that GR has to incorporate is
that gravitational field itself has energy and hence like any other energy
it must also link to gravity. That is, field has gravitational charge and
hence it is self interacting. Field energy density will go as square of
first derivative of the metric and it must be included in the Einstein
field equation. It is indeed included for the Riemann curvature involves
the second derivative and square of the first derivative. However in the
specific case of field of an isolated body, we obtain $1/r$ potential, the
same as in the Newtonian case. Where has the square of $\nabla \Phi$ ($\Phi$ denotes the gravitational potential) gone?
It turns out that its contribution has gone into curving the space,
$g_{rr}$ component of the metric being different from $1$. \par
The main point is that gravitational field equation should follow from the
curvature of spacetime and they should be second order quasilinear
differential equations (quasilinear means the highest order of derivative
must occur linearly so that the equation admits a unique solution).
Riemann curvature through the Bianchi identities leads to the Einstein
equation with the $\Lambda$ term. We should emphasize here that $\Lambda$
enters here as naturally as the stress energy tensor. It is indeed a true
new constant of the Einsteinian gravity \cite{n2}. 
It is a pertinent question to
ask, is this the most general second order quasilinear equation one can
obtain from curvature of spacetime? The answer is No. There exists a
remarkable combination of square of Riemann tensor and its contractions,
which when added to the action gives a second order quasilinear equation
involving second and fourth power of the first derivative. This is what is
the famous Gauss-Bonnet (GB) term. Thus GB term too appears naturally and
should have some non-trivial meaning.\par
However GB term is topological in $D<5$ and hence has no dynamics. It
attains dynamics in $D>4$. Note that gravity does not have its full
dynamics in $D<4$ and hence the minimum number of dimensions required for
complete description of gravitation is $4$. This self interaction of
gravity arises through square of first derivative of the metric. Self
interaction should however be iterative and hence higher order terms
should also be included. It turns out that there exists generalization of the 
GB term in higher dimensions in terms of the Lovelock Lagrangian 
which is a polynomial in the Riemann curvature. That again yields the 
quasilinear second order equation with higher powers of the first derivative. 
Thus GB and Lovelock Lagrangian represent higher order loop corrections to 
the Einstein gravity. \par
They do however make non-trivial contribution 
classically only for $D>4$ dimensions. This is rather important. If GB term 
had made a non-trivial
contribution in $4$-dimensions, it would have conflicted with the $1/r$
character of the potential because of the presence of $(\nabla\Phi)^4$
terms in the equation. The square terms (to account for contribution of gravitational field energy) were taken care of by the space curvature ($g_{rr}$ in
the metric) and now nothing more is left to accommodate the fourth (and higher) power term. However we can not tamper with the inverse square law (i.e. $1/r$ potential) which is
independently required by the Gauss law of conservation of flux. That can
not be defied at any cost. Thus it is not for nothing that the GB and its 
Lovelock generalization term
makes no contribution for $D=4$. It further carries an important message
that gravitational field cannot be kept confined to $4$-dimensions. It is
indeed a higher dimensional interaction where the higher order iterations 
attain meaning and dynamics. This is the most profound message
the GB term indicates. This is yet another independent and new motivation
for higher dimensional gravity \cite{n2}. \par
 
Self interaction is always to be evaluated iteratively. For gravity
iteration is on the curvature of spacetime. It is then not surprising that GB
term arises naturally from the one loop correction to classical gravity.
String theory should however encompass whatever is obtained by iterative
the iterative process. GB term is therefore strongly motivated by
string theoretic considerations as well.
Further GB is topological in $4$-D but in quantum considerations it
defines new vacuum state. It is quantum mechanically non-trivial. In
higher dimensions, it attains dynamics even at classical level. In the
simplest case in higher dimension it should have a classical analogue of
$4$-D quantum case. That is what indeed happens. Space of constant
curvature or equivalently conformally flat Einstein space solves the
equation with GB term with redefined vacuum. This is a general result for
all $D>4$. It is interesting to see quantum in lower dimension becoming
classical in higher dimension. \par
 
In the context of the brane bulk system we should therefore include GB
term in the bulk and see its effects on the dynamics on the brane. The
brane world gravity should thus be studied with GB term not necessarily as
correction but in its own right. It is a true description of high energy
gravity. However, for the purpose of following discussion, we shall treat
GB as a correction term. \par

The Einstein-Gauss-Bonnet action for five dimensional bulk containg a 4D brane is 
\begin{eqnarray}
S &=& \frac{1}{2 \kappa_5^2}\int d^5x\sqrt{-g}\big\lbrace {\cal R} - 2\Lambda_5
+ \alpha_{\rm GB} \lbrack {\cal R}^2 - 4{\cal R}_{AB}{\cal R}^{AB} \nonumber\\
&+& {\cal R}_{ABCD} {\cal R}^{ABCD}\rbrack \big \rbrace
+ \int d^4x\sqrt{-h} ({\cal L}_m - \lambda_b)~,
\end{eqnarray}
${\cal R}, R$ refer to the Ricci scalars in the bulk metric $g_{AB}$ and the induced metric on the brane 
$h_{AB}$; $\alpha_{rm GB}$ has dimensions of ({\em length})$^2$ and
is the Gauss-Bonnet coupling, while $\lambda_b$ is the brane tension
 and $\Lambda_5\,(<0)$ is the bulk
cosmological constant. The constant $\kappa_5$ contains the $M_5$, the 5D fundamental energy scale ($\kappa_5^2=M_5^{-3}$).
The
modified Friedman equation on the (spatially flat) brane 
may be written as~\cite{D,T,lidsey} (see also Ref\cite{g})
 \begin{eqnarray}
H^2 &=& {1\over
4\alpha_{\rm GB}}\left[(1-4\alpha_{\rm GB}\mu^2)\cosh\left({2\chi\over3}
\right)-1\right]\,,\label{mfe}\\
\label{chi} \kappa_5^2(\rho+\lambda_b) &=&
\left[{{2(1-4\alpha_{\rm GB}\mu^2)^3} \over {\alpha}_{\rm GB} }\right]^{1/2}
\sinh\chi\,,
 \end{eqnarray}
where $\chi$ is a dimensionless measure of the energy-density.
In order to regain general relativity at low energies, the
effective 4D Newton constant is defined by~\cite{T}
\begin{equation}\label{m4}
\kappa_4^2\equiv {1 \over M_p^2}= {\kappa_5^4\lambda_b\over
6(1-4\alpha_{\rm GB}\Lambda_5/9)}\,.
\end{equation}
When $\alpha_{\rm GB}=0$, we recover the RS expression. We can fine-tune
the brane tension to achieve zero cosmological constant on the
brane~\cite{T}:
\begin{equation}\label{sig}
\kappa_5^4\lambda_b^2=-4\Lambda_5+{1\over\alpha_{\rm GB}}\left[1 -\left(1+
{4\over3} \alpha_{\rm GB}\Lambda_5\right)^{\!3/2} \right].
\end{equation}
The modified Friedman equation~(\ref{mfe}), together with
Eq.~(\ref{chi}), shows that there is a characteristic GB energy
scale $M_{\rm GB}$\cite{DR} such that,
 \begin{eqnarray}
\rho\gg M_{\rm GB}^4~& \Rightarrow &~ H^2\approx \left[ {\kappa_5^2 \over
16\alpha_{\rm GB}}\, \rho \right]^{2/3}\,,\label{vhe}\\
M_{\rm GB}^4 \gg \rho\gg\lambda_b~& \Rightarrow &~ H^2\approx {\kappa_4^2
\over
6\lambda_b}\, \rho^{2}\,,\label{he}\\
\rho\ll\lambda_b~& \Rightarrow &~ H^2\approx {\kappa_4^2 \over 3}\,
\rho\,. \label{gr}
 \end{eqnarray}
 \bigskip
 \begin{figure}
 \vspace{0.5in}
 \resizebox{3.5in}{!}{\includegraphics{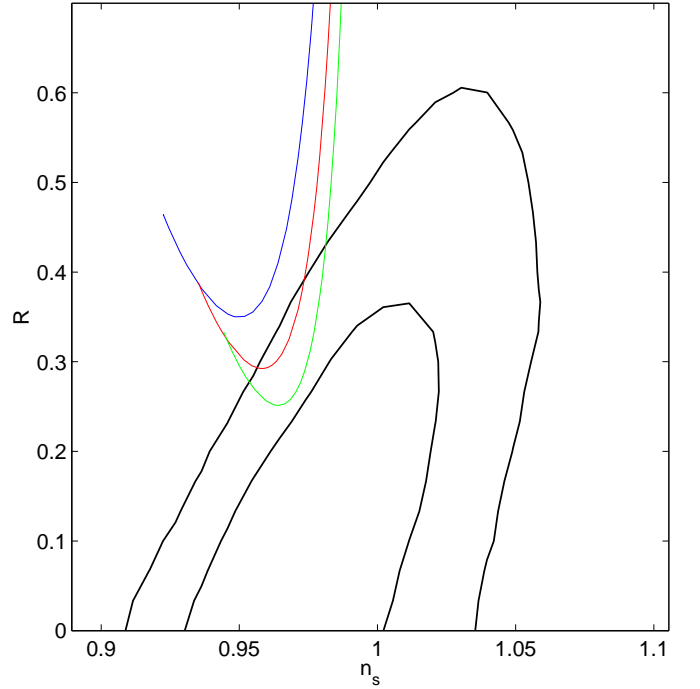}}
 \caption{ Plot of $R$ ($R \equiv 16 A_T^2/A_S^2$ $-$ according to the normalization used here\cite{ssr}) versus the spectral index $n_S$ in case of the exponential potential for ${\cal N}=50,60,70$ (from top to bottom) along
 with the $1\sigma$ and $2\sigma$ observational contours. These curves exhibit a minimum in the
 intermediate region between GB (extreme right) and the RS (extreme left) regimes. The upper limit
 on $n_S$ is dictated by the quantum gravity limit where as the lower bound is fixed by the
 requirement of ending inflation in the RS regime\cite{ssr}. For a larger
 value of the number of e-folds ${\cal N}$, more points are seen to be within the $2\sigma$ bound. Clearly, steep inflation in the deep GB regime is not favored due to the large value of $R$ inspite the spectral index being very close to $1$ there. From Sami and Sahni\cite{samiv}}
 \label{exponential.ps}
 \end{figure}
It should be noted that Hubble law acquires an unusual form for energies higher that than the
GB scale. Interestingly, for an exponential potential, the modified Eq.(\ref{vhe}) leads to 
exactly scale invariant spectrum for primordial density perturbations. Inflation continues below GB scale
and terminates in the RS regime leading to the spectral index very close to one. This is amazing
that it happens without tuning the slope of the potential. The Gauss-Bonnet inflation has
interesting consequences for steep brane world inflation (see Fig. \ref{exponential.ps} and the discussion in the next section).
\section{Summary}
In this paper we have reviewed the recent work on unification of inflation with quintessence
in the frame work of brane worlds. These models belong
to the class of {\it non-oscillatory} models in which the underlying alternative reheating mechanism
plays a crucial role. The popular reheating alternative via quantum mechanical production of
particle during inflation leads to an unacceptable relic gravity wave background which
violates the nucleo-synthesis constraint at the commencement of radiative regime. We have mentioned other alternatives to conventional (p)reheating and have shown that 'instant preheating' discovered by
Felder, Linde and Kofman is superior and best suited to brane world models of quintessential inflation. The recent measurement of CMB anisotropies
by WMAP, appears to heavily constraint these models. 
The steep brane world inflation seems to be excluded by observation in
RS scenario\cite{suji04}. As shown in Ref.\cite{ssr}, the inclusion of GB term in the bulk effects the constraints on the
inflationary
potentials and  can rescue the steep exponential potential allowing it to be compatible
with observations for a range of energy scales. The GB term leads to an increase of the spectral index $n_S$
and decrease of tensor to scalar ratio of perturbations $R$ in the intermediate region between RS and GB.
As seen from Fig.\ref{exponential.ps}, there is an intermediate region
where the steep inflation driven by exponential potential lies within $2 \sigma$ contour for  ${\cal N}=70$ .
Thus, the steep inflation in a Gauss-Bonnet braneworld appears to be
in agreement with observations.


\section{Acknowledgments}
We thank N. Mavromatos, S. Odintsov, V. Sahni and Shinji Tsujikawa for useful comments.

\end{document}